\documentclass[12pt]{article}
\usepackage{epsf}

\begin{document}

\begin{flushright}
SINP/TNP/02-11\\
UMD-PP-02-045\\
\end{flushright}

\def\n{\nu}
\def\nel{\nu^e_L}
\def\nml{\nu^\mu_L}
\def\ntl{\nu^\tau_L}
\def\ner{\nu^e_R}
\def\nmr{\nu^\mu_R}
\def\ntr{\nu^\tau_R}
\def\nsl{\nu^s_L}
\def\nsr{\nu^s_R}
\def\nlm{\nu^-_L}
\def\nlp{\nu^+_L}
\def\nrm{\nu^-_R}
\def\nrp{\nu^+_R}
\def\mp{M^\prime}
\def\mpp{M^{\prime \prime}}
\def\mppp{M^{\prime \prime \prime}}

\def\nep{\nu^\prime_e}
\def\nmp{\nu^\prime_\mu}

\newcommand{\be}{\begin{eqnarray}}
\newcommand{\ee}{\end{eqnarray}}
\newcommand{\etal}{{\it et al.}}

\def\nue{{\nu_e}}
\def\anue{{\overline{\nu_e}}}
\def\numu{{\nu_{\mu}}}
\def\anumu{\overline{\nu_{\mu}}}
\def\nutau{{\nu_{\tau}}}
\def\anutau{{\bar\nu_{\tau}}}
\def\epsi{\epsilon ~\sin 2 \theta }

\newcommand{\dm}{\mbox{$\Delta{m}^{2}$~}}
\newcommand{\st}{\mbox{$\sin^{2}\theta$~}}
\newcommand{\br}{\mbox{$^{8}{B}~$}}
\newcommand{\ber}{\mbox{$^{7}{Be}$~}}
\newcommand{\cl}{\mbox{$^{37}{Cl}$~}}
\newcommand{\ga}{\mbox{$^{71}{Ga}$~}}

\newcommand{\memu}{\mbox{$m_{e\mu}$~}}
\newcommand{\metau}{\mbox{$m_{e\tau}$~}}
\newcommand{\mmutau}{\mbox{$m_{\mu\tau}$~}}

\begin{center}
{\large \bf A MINIMAL THREE GENERATION SEESAW SCENARIO FOR LSND}
\vspace{.5in}

{\bf Biswajoy Brahmachari$^{a}$, Sandhya Choubey$^{b}$ and 
Rabindra N. Mohapatra$^{c}$\\}
 
\vskip .5cm
 
(a) Theoretical Physics Group, Saha Institute of Nuclear Physics,\\
1/AF Bidhannagar, Kolkata 700064, INDIA \\
\vskip .5cm    

(b) Department of Physics and Astronomy, University of Southampton, \\
Highfield, Southampton S017 1BJ, UK
\vskip .5 cm    

(c) Department of Physics and Astronomy, University of Maryland,\\
 College Park, MD-20742, USA\\

\end{center}

\vskip 3cm    

\begin{center}
\underbar{Abstract}
\end{center}
We show that in the minimal three generation seesaw models for neutrinos,
the presence of leptonic $(L_e+L_{\mu}-L_{\tau})\times S_2$ symmetry leads
to one of the right handed neutrinos remaining massless. This state can
then be identified with the sterile neutrino required for a
simultaneous understanding of solar, atmospheric and LSND observations.
We present a gauge model where the presence of higher dimensional
operators originating from Planck scale physics lead to a realistic 2+2
mixed scenario that fits all
oscillation data. The model predicts a range for the mixing angle $U_{e3}$
and an effective mass for neutrinos emitted in tritium decay, which can be
used to test this model.
\newpage

\section{Introduction}

The evidence for neutrino masses and mixings now appears to be quite solid
from the positive results for neutrino oscillations from solar
neutrino data in seven different experiments, Chlorine, Kamiokande,
Super-Kamiokande, SAGE, GALLEX, GNO, SNO\cite{ref1} and atmospheric
neutrino data from Super-Kamiokande, IMB, Soudan and MACRO
experiments\cite{ref2}. The third piece of evidence is from the Los Alamos
LSND experiment that shows an oscillation from the muon neutrino to 
the electron neutrino (from $\anumu$ to $\anue$)\cite{lsnd}. The
KARMEN\cite{karmen} experiment which looked for $\anumu$ to $\anue$
oscillation did not find any evidence for it and eliminated a large
fraction of the parameter space allowed by LSND. It is hoped
that the Mini-BOONE experiment at FERMILAB\cite{mb} will settle the issue
in near future.

Theoretical analyses have made it clear that a simultaneous explanation
of all the neutrino oscillation observations requires the existence of an
ultralight sterile neutrino in addition to the three known active
neutrinos $(\nu_e, \nu_{\mu}, \nu_{\tau})$\cite{cald}. There are two mass
patterns that are currently under active discussion; the 2+2
scheme\cite{cald} and 3+1 scheme\cite{3p1}. The 2+2 scheme has two mass 
eigenstate neutrinos with mass around an eV and a lighter pair with mass 
near $10^{-3}$ eV, with the latter pair explaining the solar neutrino 
data, the former explaining the
atmospheric neutrino data and the gap between the two pairs explaining
the LSND results. Recent SNO data disfavors the original version
of the 2+2 model where all the missing solar $\nu_e$'s are converted via
a small angle MSW mechanism only to the sterile neutrinos. The situation 
where only a fraction of the missing $\nu_e$s convert to $\nu_s$s and the 
rest to active ones has been studied in detail\cite{concha} and shown to be
quite viable at the moment. The 3+1 picture on the other hand is severely
constrained by the known accelerator data but is also viable for certain
values for \dm.

If indeed the sterile neutrino turns out to be required, an important
theoretical challenge is to seek an understanding of such tiny mass of a
particle which is neutral under the standard model forces. This is in
clear distinction from the case for the active neutrinos for whom the
small mass has a very convincing explanation in terms of the seesaw
mechanism\cite{seesaw} that, at the basic level requires only that three
singlet right handed neutrinos be introduced with masses close to the GUT
scale. The conventional seesaw mechanism makes the three right handed
neutrinos superheavy, leaving the three active neutrinos with a very light
mass, as seems to be case. In addition to this appealing feature, this
minimal  seesaw
model also restores quark lepton symmetry into the fundamental fermions
of nature, making this the standard paradigm of neutrino mass physics. 
Generic attempts to understand the ultralight sterile
neutrino within the seesaw framework involve the introduction of a new
singlet fermion into the theory beyond the three right handed neutrinos
just discussed and use extra symmetries to protect it from being
superheavy. While there exist perfectly acceptable, interesting and
technically natural scenarios for the sterile neutrino with extra
singlet fermions\cite{models}, these extra assumptions have caused
skepticism regarding the existence of the sterile neutrino and in turn
about the LSND results.

An interesting theoretical challenge is therefore to explore whether
within the
minimal quark lepton symmetric seesaw framework, one can have an
ultralight 
sterile neutrino without the need for additional singlet neutrinos,
outside the quark-lepton symmetric framework and
fit all observations including LSND. Such a project is apriori not
implausible
in view of the fact that the right handed neutrinos already present in the
quark-lepton symmetric framework have all the properties
of a sterile neutrino, except the mass.
In a recent paper\cite{ss-sterile} it was shown that if there exist certain
leptonic symmetries in the seesaw model, they can make one of
the right handed neutrinos massless in the tree
approximation of a renormalizable theory
making it possible to identify it with the needed ultralight sterile
neutrino. Using this idea for the case of an
$(L_e-L_{\mu}-L_{\tau})\times S^{\mu\tau}$
symmetry, (where $S^{\mu\tau}$ is the interchange symmetry between
$\nu_{\mu R}$ and $\nu_{\tau R}$), it was shown in \cite{ss-sterile} that
the 3+1 scenario for the LSND observations can be reproduced within the
minimal seesaw picture.

In this letter, we show that when the leptonic symmetry is instead chosen
to be $(L_e+L_{\mu}-L_{\tau})\times S_2$, the suggestion of
ref.\cite{ss-sterile} can be extended to obtain the 2+2 explanation
of LSND with one of the right handed neutrinos of the seesaw picture
playing the role of the sterile neutrino. The mixing of the sterile
neutrino to the active neutrinos arises from Planck scale suppressed
higher dimensional operators. A consistent picture with $10^{13}$ GeV for
the seesaw scale and $10^6$ GeV for the breaking of the leptonic
symmetries seems to emerge and leads to mass matrix ansatz for the 2+2
case recently proposed in \cite{babu}. Experimental tests of this idea are
proposed.

\section{Seesaw with a leptonic symmetry and right handed neutrino as an
ultralight sterile neutrino} 
In this section, we discuss how imposing $(L_e+L_{\mu}-L_{\tau})\times
S_{e\mu}$ in a seesaw framework leads to a massless right handed neutrino
which can play the role of the sterile neutrino. 
We will work with the gauge group $SU(2)_L\times U(1)_{I_{3R}}\times
U(1)_{B-L}$ with particle assignment as in Table I.
\begin{table}
\begin{center}
\begin{tabular}{|c|c|c|c|c|}\hline
& $SU(2)_L \times U(1)_{I_{3R}} \times U(1)_{B-L}$ & $SU(2)_L\times
U_{Y}(1)$ 
& $L_e+L_\mu-L_\tau$ & $S_{2R}^{e\mu}$ \\\hline

$\nu_{-R}$ &(1,1/2,-1) & (1,0) & 1 & -1 \\
$\nu_{+R}$ &(1,1/2,-1)& (1,0) &  1 & 1 \\
$\nu_{\tau R}$ &(1, 1/2,-1)& (1,0) & -1 & 1 \\
\hline
$\Delta$ &(1,-1,+2)& (1,0) & 0 & 1\\
$\phi$ &(2, 1/2,0)& (2,1/2) & 0 & 1 \\
$\sigma_2$ & (1,0,0) & (1,0) & -2 & -1 \\
$\sigma_0$ & (1,0,0) & (1,0) & 0 & -1 \\\hline
\end{tabular}
\end{center}
\caption[]{Relevent right handed fermion and 
scalar fields and their transformation
properties. Here we have defined $Y=I_{3R}+{B-L \over 2 }$\label{tab1}}
\end{table}
We call this the minimal seesaw
picture since all the fermions in our model are needed in the conventional
seesaw explanation of the small masses of the neutrinos.
The Higgs field $\Delta^0$ transforming as $ (1, -1, +2)$ under the 
gauge group couples to the
right handed neutrinos 
($\nu_{eR}, \nu_{\mu R}, \nu_{\tau R}$) with couplings typically of the
form $\nu_R\nu_R \Delta$. After spontaneous symmetry breaking, the field
$\Delta$ acquires a vev i.e. $<\Delta^0> = v_R$ and breaks the gauge
symmetry to the standard model. This gives mass to the right handed
neutrinos and is the seesaw scale. The Higgs field $\phi (2, 1/2, 0)$,
whose weak scale vev
breaks the standard model gauge group, gives Dirac mass to the neutrinos. 
If we denote
the mass matrix for the right handed neutrinos as $M_R$, then the complete
$6\times 6$ mass matrix involving the Dirac mass and Majorana mass for the
neutrinos can be written as
\begin{eqnarray}
{\cal M}_{LR}~=~ \left(\begin{array}{cc} 0 & M_D \\ M^T_D &
M_R\end{array}\right).
\end{eqnarray}
When none of the eigenvalues of $M_R$ vanishes, one can obtain the mass
matrix for
the light neutrinos as
\begin{eqnarray}
{\cal M}_{\nu}~=~ - M^T_DM^{-1}_RM_D
\end{eqnarray}
This is the so-called type I seesaw formula.
On the other hand when $M_R$ and $M_D$ matrices have zero eigenvalues, 
one must ``take them out'' of the
matrix before using the seesaw formula. As was noted in \cite{ss-sterile},
this turns out to be the case when there are leptonic symmetries. To
obtain the 2+2 scenario, we will consider the leptonic symmetry to be
$(L_e+L_\mu-L_\tau)\times S_{2R}^{e\mu}$.
At the beginning of our analysis we go to a basis where charged 
lepton mass matrix is diagonal where leptonic
mixing matrix is simply neutrino mixing matrix.
In this basis we identify the leptonic flavor
indices $e,\mu,\tau$. After that the renormalizable Lagrangian invariant 
under $(L_e+L_\mu-L_\tau) \times S_{2R}^{e\mu}$ symmetry has the form
\begin{eqnarray}
{\cal L}^R_Y &=& h_e ~\bar{L}_e\phi e_R 
+h_{\mu} ~\bar{L}_{\mu}\phi \mu_R 
+ h_{\tau} ~\bar{L}_{\tau}\phi \tau_R \nonumber\\  
&& +h_1 ~\bar{L}_e\tilde{\phi} ~(\nu_{eR}+\nu_{\mu R})  
+h_2 ~\bar{L}_{\mu}\tilde{\phi} ~(\nu_{eR}+\nu_{\mu R})  
+h_3 ~\bar{L}_{\tau}\tilde{\phi}\nu_{\tau R} \nonumber\\ 
&& + f ~(\nu_{\mu R} +\nu_{e R})\Delta~ \nu_{\tau R}           
\end{eqnarray}
Spontaneous symmetry breaking then leads to Dirac and Majorana type
mass matrices of the form, 
\be
M_D = \bordermatrix {&\ner&\nmr&\ntr \cr
\hline
&&&\cr
\overline{\nel} & k & k & 0\cr
\overline{\nml}&k^\prime & k^\prime & 0 \cr
\overline{\ntl} & 0 & 0 & m_{33}\cr},~~~~
M_R=\bordermatrix
{& \ner & \nmr & \ntr \cr
\hline
&&&\cr
\ner   & 0 & 0 & M \cr
\nmr & 0 & 0 & M  \cr
\ntr & M & M & 0  \cr}
\ee
The $6\times 6$ mass matrix\cite{ss-sterile} would then take the form
\be
M = \bordermatrix
{&\nel  & \nml & \ntl & \ner & \nmr & \ntr \cr
\hline
&&&&&&\cr
\nel & 0 & 0 & 0 & k & k & 0 \cr
\nml & 0 & 0 & 0 & k^\prime & k^\prime & 0 \cr
\ntl & 0 & 0 & 0 & 0 & 0 & m_{33} \cr
\ner & k & k^\prime & 0 &  0 & 0 & M \cr
\nmr &  k & k^\prime & 0 &  0 & 0 & M \cr
\ntr & 0 & 0 & m_{33} &  M & M & 0 \cr}
\ee
In the rotated basis it is
\be
M = \bordermatrix
{& \nlm & \nlp & \ntl & \nrm & \nrp & \ntr \cr
\hline
&&&&&&\cr
\nlm &0&0&0&0&0&0 \cr
\nlp &0&0&0&0&\xi&0 \cr
\ntl &0&0&0&0&0&m_{33} \cr
\nrm &0&0&0&0&0&0 \cr
\nrp &0&\xi&0&0&0&\sqrt{2}M \cr
\ntr &0&0&m_{33}&0&\sqrt{2}M&0 \cr}
\ee
where 
\begin{eqnarray}
\nu^-_L &=& \frac{k^\prime \nel - k \nml}{\sqrt{{k^\prime}^2+k^2}}
\label{rot1}
\\
\nu^+_L &=& \frac{k \nel+k^\prime \nml}{\sqrt{{k^\prime}^2+k^2}}
\label{rot2}
\\
\nu^\pm_R& =& \frac{\ner \pm \nmr}{\sqrt{2}}
\end{eqnarray}
and $\xi=\frac{2\sqrt{2}kk^\prime}{\sqrt{{k^\prime}^2+k^2}}$. 
We 
next employ the see-saw mechanism to generate the light Majorana 
neutrino mass matrix,
\begin{eqnarray}
\bordermatrix
{& \nlp & \ntl & \nlm \cr
\hline
&&&\cr
\nlp   & 0 & m & 0  \cr
\ntl & m & 0 & 0  \cr
\nlm & 0 & 0 & 0 \cr}~~~{\rm where}~~~ m={m_{33} \xi \over \sqrt{2} M}
\label{atmlsnd}
\end{eqnarray}
In the
limit $k=k^\prime$ we have $\xi= 2 k$ and $m={\sqrt{2} k m_{33} \over M}$
\cite{ss-sterile}, however because $S_{2L}$ symmetry is absent we have 
the freedom to
choose $k \ne k^\prime$ which will help to achieve small mixing in 
the $\nue-\numu$ sector compatible to LSND.

The challenge after this is to induce large left right mixing.
There are higher dimensional operators in this scenario which are 
sufficient to induce left-right mixing after the see-saw mechanism
as will be described below. We can write down following higher dimensional 
operators
\begin{eqnarray}
&& {f_1 \over M_p}~ \overline{L_e} \nu_{-R} \langle \sigma_0 \rangle 
\langle \tilde{\phi} \rangle \\
&& {f_2 \over M_p}~ \overline{L_\mu} \nu_{-R} \langle \sigma_0 \rangle 
\langle \tilde{\phi} \rangle \\
&& {f_3 \over M_p }~ \overline{L_\tau} \nu_{-R} \langle \sigma_2 \rangle
\langle \tilde{\phi} \rangle \\
&& {f_4 \over M^2_p}~\nu_{-R} \nu_{-R} \langle \Delta \rangle 
\langle \sigma_2 \rangle \langle \sigma_0 \rangle 
\end{eqnarray}

Now the $4\times 4$ mass matrix in low energy can be written in terms
of rotated low energy fields. We identify $\nlp \equiv \nep, \nlm \equiv
\nmp,  \nu_{-R} \equiv \nu_s$
\begin{equation}
M=\bordermatrix{ & \nep & \nmp & \ntl & \nu_s \cr
\hline
\nep & 0 & 0 & m & m_1 \cr
\nmp & 0 & 0 & 0 & m_2 \cr
\ntl & m & 0 & 0 & m_3 \cr
\nu_s & m_1 & m_2 & m_3 & \delta \cr
 } 
\label{mat1}
\end{equation}
\vskip .5cm
\noindent \underbar{CASE 1}
\vskip .5cm
We notice that in the limit $m \rightarrow 0, \delta \rightarrow 0$
the eigenvalues are
\begin{equation}
0, 0, \pm M {\rm ~~~where~~~}M=\sqrt{m^2_1  + m^2_2  + m^2_3}
\end{equation}
Thus LSND mass splitting squared can be of order $M^2$. Introduction of
non-zero $m$ and $\delta$ will introduce $\Delta m^2_\odot$ and
$\Delta m^2_{atm}$. To leading order the eigenvectors are,
\begin{eqnarray}
&& {\rm Eigenvalue~0}~~~ 	e_1  = \pmatrix{            
	- m_3 /\sqrt{m^2_1 + m^2_3} \cr                
	0 \cr             
	m_1 /\sqrt{m^2_1 + m^2_3} \cr                
	0} ~~~;~~~ 
e_2 = \pmatrix{            
	-m_1 m_2 /M \sqrt{m^2_1 + m^2_3} \cr             
	\sqrt{(m^2_1+m^2_3)}/M   \cr            
	-m_2 m_3 /M \sqrt{m^2_1 + m^2_3} \cr                           
	0}  
\nonumber\\          
&& {\rm Eigenvalue ~\pm M}~~~   e_3  =  \pmatrix{
           -m_1 /\sqrt{2} M\cr
           -m_2 /\sqrt{2} M\cr
           -m_3 /\sqrt{2} M\cr
            1/ \sqrt{2}}~~~;~~~ 
e_4  = \pmatrix{
           m_1 /\sqrt{2} M\cr
           m_2 /\sqrt{2} M\cr
           m_3 /\sqrt{2} M\cr
            1/ \sqrt{2}} \label{mixing}            
\end{eqnarray} 
Depending on the hierarchies between $m_1,m_2,m_3,\delta$ and $m$ various
scenarios can emerge from general expressions given in Eqn. (\ref{mixing}). 
Let us consider a simple possibility. Taking $m_2 >> m_1,m_3, 
m, \delta$, and $m_1 \approx m_3$ we have $M=m_2$. Eigenvectors reduce to 
\begin{eqnarray}
&&  
e_1  = \pmatrix{            
	- 1 /\sqrt{2} \cr                
	0 \cr             
	1 /\sqrt{2} \cr                
	0} ~~~;~~~ 
e_2 = \pmatrix{            
	-1 /\sqrt{2} \cr             
	0   \cr            
	-1 / \sqrt{2} \cr                           
	0} ~~~;~~~ 
e_3  =  \pmatrix{
           0 \cr
           -1/\sqrt{2} \cr
           0 \cr
            1/ \sqrt{2}}~~~;~~~ 
e_4  = \pmatrix{
           0 \cr
           1 /\sqrt{2} \cr
           0 \cr
            1/ \sqrt{2}} \label{mixing1}            
\end{eqnarray} 
This means that the massive pair is constituted mainly of $\nu^\prime_\mu$
and $\nu_s$, whereas the light pair is constituted mainly of
$\nu^\prime_e$ and $\nu_\tau$ and each pair has large mixing among them. 
Now take $k >> k^\prime$ in Eqn. (\ref{rot1}) and Eqn. (\ref{rot2}).
Then $\nu^\prime_e \approx \nel$ and $\nu^\prime_\mu \approx \nml$. 
We must introduce non-zero $m$, $\delta$ $m_1$ and $m_2$ in such a way 
that the two pairs mix among each other to explain the LSND result.

This can be done numerically for example if we take the parameter set,
$m=0.04, m_1=0.05, m_2=1, m_3=0.06, \delta=0.02$ eVs. We get,
$
\Delta m^2_\odot=2 \times 10^{-5},
\Delta m^2_{atm}=4 \times 10^{-2},
\Delta m^2_{LSND}=0.98
$ eV$^2$.
\begin{eqnarray}
&&  
e_1  = \pmatrix
{0.707 \cr  -0.007 \cr -0.706 \cr 0}
e_2 = \pmatrix 
{0.705 \cr  -0.077 \cr  0.704 \cr -0.003}
e_3  =  \pmatrix
{0.033 \cr  0.708 \cr 0.041 \cr 0.703}
e_4  = \pmatrix
{-0.036 \cr -0.701 \cr -0.044 \cr 0.710}
\label{mixing2}            
\end{eqnarray} 
We note that now we have introduced small $\nu_e-\nu_\mu$ mixing
even though the primary mixing pattern is governed 
by Eqn. (\ref{mixing1}). We can now calculate 
P($\anue \rightarrow \anumu$) for LSND experiment to be  
\begin{equation}
P_{\anue \rightarrow \anumu}^{LSND} 
= 4 |(U_{e3}^* U_{\mu 3} + U_{e4}^* U_{\mu 4})|^2
\sin^2 (\frac{1.27 \Delta m^2_{LSND} L}{E})
\end{equation}
On the other hand, we have,
\begin{equation}
P_{\anue \rightarrow \anumu}^{LSND} = \sin^2 2\theta_{LSND} 
\sin^2 (\frac{1.27 \Delta m^2_{LSND} L}{E})
\end{equation}
Combining these two equations we have
\begin{equation}
\sin^2 2\theta_{LSND}=0.01
\end{equation}
Thus we see that we get $\Delta m^2$ as well as $\sin^2 2 \theta$
compatible with LSND experiment\cite{lsnd}. Whereas we have already
got large(near-maximal) $\nu_e \leftrightarrow \nu_\tau$ and 
large(near-maximal) $\nu_\mu \leftrightarrow
\nu_s$ mixing to account for solar and atmospheric neutrino deficits.
However in two-generation analysis of the atmospheric neutrino data
the {\it pure} $\nu_\mu \leftrightarrow \nu_s$ oscillation scenario
is seen to be highly disfavored, as it fails to reproduce the
correct zenith angle distribution of the data, the effect being more
for the higher energy neutrinos. The sterile option is also
in contradiction with the neutral current sample of the SK atmspheric
neutrino data \cite{sksterile}. But from the four-neutrino
global analysis of the solar and the atmospheric neutrino data
taken together \cite{concha}
this scenario is not ruled out, although disfavored \cite{concha}.
The other drawback of this mass texture is that it predicts
near-maximal mixing for the solar $\nu_e \leftrightarrow \nu_\tau$
oscillations. However, after the inclusion of the SNO data,
maximal mixing in the LMA region is seen to be disfavored
in most analyses of the solar neutrino problem \cite{postsno}.

\vskip .5cm
\noindent \underbar{CASE 2}
\vskip .5cm
Let us apply the inverse of the 
rotation given in Eqn (\ref{rot1}) and Eqn (\ref{rot2}) on the basis vectors
$(\nu^\prime_e,\nu^\prime_\mu)$. Then matrix in 
Eqn. (\ref{mat1}) can be expressed as,
\begin{equation}
M=\bordermatrix{ & \nel & \nml & \ntl & \nu_s \cr
\hline
\nel & 0 & 0 & \epsilon_2 & \epsilon_1 \cr
\nml & 0 & 0 & 1 & a \cr
\ntl & \epsilon_2 & 1 & 0 & b \cr
\nu_s & \epsilon_1 & a & b & \delta \cr
 } m_0
\end{equation}
For small b this mass matrix has been studied in Ref. \cite{babu} 
using a different symmetry than the one discussed above. 
It was shown in detail that it is suitable for the mixed $2+2$ scenario.
A similar matrix was also obtained under radiative scheme. However the 
44 diagonal element vanish genericaly in simple radiative schemes\cite{anjan}.

\section{Predictions of the Model}
With the range of values for the parameter set needed to explain the 
global solar, atmospheric and the accelerator/reactor data including 
LSND, our four-generation model makes definite predictions for 
the effective mass of the neutrinos in beta decay and neutrinoless 
double beta decay processes. For the neutrinoless double beta decay 
the ``effective Majorana mass parameter'' is given by
\begin{eqnarray}
|<m>| &=& |\sum_{i=1,4}m_i U_{ei}^2|
\\ \nonumber
&=& m_{ee}
\end{eqnarray}
the $11$ element of the mass matrix. Since this element is identically 
zero for the model under consideration, our model predicts no 
neutrinoless double beta decay. 
In a recent paper \cite{klapdor} 
has claimed of a  positive evidence for neutrinoless double beta decay, 
based on the reanalysis of the data from the Heidelberg-Moscow experiment. 
A best-fit Majorana mass of 0.39 eV for the neutrinos at 95 \% 
confidence level was reported. However there are 
proposals for new generation neutrinoless double beta decay 
experiments which are expected to have better sensitivities
so they can resolve this issue at higher confidence levels. These 
include NEMO3 \cite{nemo3}, CUORE \cite{cuore} and GENIUS \cite{genius}, 
which hopes to probe neutrino masses down to $\sim 10^{-2}$ eV.
This is a good way to falsify the scenario we are presenting here. 

The beta decay experiments look for evidence of neutrino mass by 
observing the end point spectrum of $^3H$ beta decay. The best bound 
available to date is \cite{mainz,troitsk} 
\begin{eqnarray}
m_{\nu_e} <  2.2 ~{\rm eV ~ at~95\%~C.L.}
\end{eqnarray}
In terms of the mass and mixing angles the quantity probed by the 
beta decay experiments is
\begin{eqnarray}
m_{\nu_e}^2 = \sum_{i=1,4}m_i^2 |U_{ei}|^2 
\end{eqnarray}
For the CASE 2 of our model we have to linear order in $\delta$, 
$\epsilon_1$ and $\epsilon_2$
\begin{eqnarray}
m_{\nu_e}^2 \approx \frac{m_0^2}{(1+a^2)} 
             \left[(\epsilon_1 - a \epsilon_2)^2 + 
              (a \epsilon_1 + \epsilon_2)^2 \right]
\end{eqnarray} 
With the range of values for the parameters $m_0^2$, 
$a$, $\epsilon_1$ and $\epsilon_2$ required to satisfy the global 
neutrino oscillation data \cite{babu} we expect the value of 
$m_{\nu_e} \sim 0.03$ for $\Delta m^2_{LSND} \sim 1$ eV$^2$. 
The forthcoming KATRIN experiment is expected to have a 
sensitivity upto $m_{\nu_e} \sim 0.3$ eV \cite{katrin}. Thus we expect 
no signal for $m_{\nu_e}$ in KATRIN from our model. This is
another way to test our model.

The other intriguing quantity that needs precise measurements is 
$|U_{e3}|^2 + |U_{e4}|^2$ which will give the fraction of $\nu_e$ mixed in 
the upper doublet. For the CASE 2 upto linear terms in $\delta$, 
$\epsilon_1$ and $\epsilon_2$ we have
\begin{eqnarray}
|U_{e3}|^2 + |U_{e4}|^2 \approx \left( \frac{a \epsilon_1 + \epsilon_2}{1+a^2}
\right )^2
\end{eqnarray} 
while
\begin{eqnarray}
4|U_{e3}^* U_{\mu3} + U_{e4}^*U_{\mu4}|^2 &\approx& 4
\left( \frac{a \epsilon_1 + \epsilon_2}{1+a^2} \right )^2
\\ \nonumber
&\approx& \sin^2 2\theta_{LSND}
\end{eqnarray} 
Thus for the CASE 2 of our model 
$|U_{e3}|^2 + |U_{e4}|^2 \approx (\sin^2 2\theta_{LSND})/4 \approx 0.0007$ 
much more stringent than the current bound from the short baseline 
terrestrial experiments (BUGEY\cite{bugey}) which is 
\begin{eqnarray}
|U_{e3}|^2 + |U_{e4}|^2 \stackrel{<}{\sim} 0.01
\end{eqnarray} 
Values of $|U_{e3}|^2 + |U_{e4}|^2$ as low as 0.0007 would be 
difficult to detect with usual laboratory beams. However with 
proposals of high intensity neutrino beams from neutrino factories,  
it would be interesting to see if one can probe to such accuracy.

\section{Comments and conclusion}

A few comments are now in order: the first theoretical comment is on the
choice of the gauge group: while we have illustrated our basic idea in the
context of an $SU(2)_L\times U(1)_{I_{3R}}\times U(1)_{B-L}$ gauge group,
it could easily be implemented in the context of the standard model gauge
group as well. However, in this case the meaning of the seesaw scale will
remain a mystery. One could also use the left-right symmetric gauge group
$SU(2)_L\times SU(2)_R\times U(1)_{B-L}$\cite{moh}. In this case since the
right handed neutrino and the right handed charged leptons are in the same
multiplet, $S_{e\mu}$ symmetry leads to electron being mass less prior to
symmetry breaking. One can however extend the model to make the electron
massive. Quark lepton unification of these models is nontrivial due to
absence of any analog of $L_e+L_{\mu}-L_{\tau}$ symmetry in the quark
sector.

The breaking of the leptonic symmetries can be made soft by adding to the
Lagrangian terms such as $(\sigma^{-}_2)^2$ so that there are no massless
scalar bosons.

In conclusion, we have presented a minimal seesaw model using three active
and three right handed neutrinos (as would be suggested in a quark-lepton
symmetric theory), which can lead to an ultralight sterile neutrino. The
sterile neutrino in this case is none other than the right handed
neutrino. Spontaneous breaking of the leptonic symmetries lead to mixing
of the sterile neutrino with the the active neutrinos resulting in a 2+2 
hybrid scenario for LSND, which provides a viable description of neutrino
oscillation data. The model predicts an effective $U_{e3}\simeq 0.025 $
and an effective end point mass in tritium beta decay $\simeq 0.03$.

The work of RNM is supported by the National Science Foundation
Grant No. PHY-0099544. B.B and S.C would like to thank the organizers
of WHEPP-7 where a part of this work was done. B.B would also like
to thank Amitava Raychaudhuri and Probir Roy for discussions on
Ref. \cite{ss-sterile}.

\end{document}